

\def\gl{\lambda}
\def\p{\partial}
\def\phip{\p_+\phi}
\def\phim{\p_-\phi}
\def\phipm{\p_+\p_-\phi}
\def\rhop{\p_+\rho}
\def\rhom{\p_-\rho}
\def\rhopm{\p_+\p_-\rho}

\rightline {SU-ITP-92-4}
\rightline {January 1992}
\bigskip\bigskip
\title{Black Hole Evaporation in 1+1 Dimensions}

\vfill
\author{Jorge G. Russo, Leonard Susskind and L\'arus Thorlacius}
\bigskip
\address{ Department of Physics \break Stanford University,
            Stanford, CA 94305}
\vfill

\abstract
\singlespace

The formation and quantum mechanical evaporation of
black holes in two spacetime dimensions
can be studied using effective classical field equations,
recently introduced by Callan {\it et al.}
We find that gravitational collapse always leads
to a curvature singularity, according to these equations, and that
the region where the quantum corrections introduced by Callan
{\it et al.} could be expected to
dominate is on the unphysical side of the singularity.
The model can be successfully applied to study the back-reaction
of Hawking radiation on the geometry of large mass black holes,
but the description breaks down before the evaporation is complete.

\vfill\endpage

\REF\hawi{S.~W.~ Hawking
\journal Comm .Math. Phys. & 43 (75) 199.}
\REF\hawii{S.~W.~Hawking
\journal Phys. Rev. & D14 (76) 2460.}
\REF\thooft{G. 't Hooft
\journal Nucl. Phys. & B335 (90) 138, and references therein.}
\REF\ac{Y.~Aharonov, A.~Casher and S.~Nussinov
\journal Phys. Lett. & 191B (87) 51.}
\REF\cghs {C.~G.~Callan, S.~B.~Giddings, J.~A.~Harvey and
A.~Strominger, {\it Evanescent black holes}, preprint UCSB-TH-91-54,
EFI-91-67, PUPT-1294, November 1991.}
\REF\russo{J.~G.~Russo and A.~A.~Tseytlin, {\it Scalar-Tensor
Quantum Gravity in Two Dimensions}, Stanford University preprint,
SU-ITP-92-2, January 1992.}
\REF\wald{R.~Wald, {\it General Relativity}, The University of
Chicago Press, 1984.}
\REF\pol{A.~M.~Polyakov
\journal Phys. Lett. & 163B (81) 207.}
\REF\witten{E.~Witten
\journal Phys. Rev. & D44 (91) 314.}

\noindent I. \ \
The discovery of Hawking that black holes emit apparently thermal
radiation raised serious questions of principle concerning the fate
of quantum information in gravitational collapse [\hawi].
Hawking himself suggested that the loss of information behind the
horizon eventually leads to a breakdown of quantum coherence
when the evaporation of the black hole is complete [\hawii].
A very different viewpoint, advocated by 't Hooft, is that quantum
corrections might transfer the information about the quantum state
of the infalling matter to the outgoing radiation in such a way
that the entire process of black hole formation and subsequent
evaporation is described by a unitary $S$-matrix [\thooft].
Yet another possibility is that the black hole does not completely
evaporate and that a Planck scale remnant carries off the
information [\ac].
To resolve these issues it will be necessary to gain some
understanding of how the Hawking radiation reacts back on the
black hole geometry.  Thus far a satisfactory description of black
hole evolution has eluded theorists, but
recent studies of black holes in 1+1 dimensions have raised hopes that
the problem might be understood in a very simplified context.
In particular, Callan, Giddings, Harvey and Strominger (CGHS) have
studied a 1+1 dimensional system, which classically has black holes
[\cghs].
They considered a renormalizable theory of two-dimensional gravity
coupled to a dilaton and a large number, $N$, of conformal
matter fields.  They found exact classical solutions of this theory
where black holes are formed in gravitational collapse.
They then added to the classical action a Liouville term which
accounts for the one-loop corrections due to the $N$ matter fields.
Because the Liouville term dominates the classical action in the
region of strong coupling the authors concluded that the usual
black hole singularity would be removed and replaced by flat
spacetime.  Without a singularity quantum information would be
guaranteed to be conserved.  Callan {\it et al.} also suggested a
large-$N$ limit to black hole physics, which might ultimately be
used as a starting point for a systematic expansion.

In this paper we analyze this theory further and show that these
hopes are not realized.   Any solution of the CGHS-equations, which
asymptotically in the past matches onto infalling matter, will produce
a serious curvature singularity whose location moves further and
further into the classical large distance region as $N$ increases.
The solution cannot be continued past the singularity into the region
studied by CGHS.
Nevertheless, for finite $N$, the model does shed some light on the
quantum evaporation of large black holes, with mass much larger than
the scale set by $N$.  In particular the solutions to the
CGHS-equations exhibit an apparent horizon which recedes due to an
outgoing flux of Hawking radiation.
The model is useful until the black hole
evaporates to a mass of order the scale set by $N$, where the apparent
horizon approaches the singularity.

\vfill
\break
\noindent
II.  \   \
The fields in the model introduced by Callan {\it et al.} are
the two-dimensional metric $g_{\mu\nu}$, a dilaton $\phi$, and
a set of $N$ massless minimally coupled scalar fields.
The classical Lagrangian is taken to be
$$
{\cal L}={1\over 2\pi} \sqrt {-g} \bigl[
  e^{-2\phi}(R+4(\nabla           \phi )^2
+4\gl ^2)-{1\over 2}\sum_{i=1}^N (\nabla f_i)^2 \bigr]\ ,
\eqn\lagr
$$
It is very similar to the target space Lagrangian of $c{=}1$
non-critical string theory, but we are simply interested in it as a
renormalizable toy model of gravity coupled to matter [\russo].  The
arbitrary parameter $\gl$ defines a mass scale.

It is worth comparing ${\cal L}$ with a model obtained by
considering spherically symmetric configurations in
four-dimensional gravity.
Let us call the 4 coordinates $x^0, x^1, \theta,\phi$ and
consider metrics of the form
$$
ds^2= g_{\mu\nu} (x^0,x^1)\, dx^\mu dx^\nu + r^2(x^0,x^1)\,
d\Omega^2  \ ,\ \ \eqn\metric
$$
where $d\Omega ^2$ is the volume element of a unit 2-sphere.
Defining $\phi \equiv -\log r$
the dimensional reduction of the Einstein-Hilbert Lagrangian is
$$
{\cal L}_{{\rm EH}}={1\over 2\pi} \sqrt {-g}
 e^{-2\phi} \bigl(R+2(\nabla \phi )^2
+2e^{2\phi } \bigr)\ .
\eqn\einst
$$
The similarity with \lagr\ is apparent and it can be useful
to think of $e^{-\phi}$ as the radius of a transverse sphere
at $x^0,x^1$.


Following reference [\cghs] we work in conformal gauge
$$\eqalign{
g_{+-} = g_{-+} =& -{1\over 2} e^{2\rho} \, ,  \cr
g_{--} = g_{++} =& \> 0 \, ,                      \cr}
\eqn\cgauge
$$
with light-cone coordinates $x^{\pm} = x^0 \pm x^1$.  The fields
in the theory are then $f_i,\phi$ and $\rho$, and their classical
equations of motion are,
$$\eqalign{
\p_+\p_- f_i =&\, 0\,,   \cr
2 \phipm - 2 \phip \phim -{\gl^2\over 2} e^{2\rho} =&\,\rhopm\,,\cr
\phipm - 2\phip\phim - {\gl^2\over 2} e^{2\rho} =&\, 0 \,, \cr}
\eqn\cgeom
$$
respectively.  In addition, one must impose as constraints the
equations of motion corresponding to the metric components, which
are set to zero in this gauge:
$$\eqalign{
e^{-2\phi}(\p_+^2\phi -2 \rhop\phip ) =&
   {1\over 4} \sum_{i=1}^N \p_+f_i \p_+f_i \, , \cr
e^{-2\phi}(\p_-^2\phi -2 \rhom\phim ) =&
   {1\over 4} \sum_{i=1}^N \p_-f_i \p_-f_i \, . \cr}
\eqn\constr
$$

A number of exact solutions of these equations can easily be found
[\cghs].  The simplest one is the vacuum solution
$$\eqalign{
f_i =&\> 0 \,,  \cr
e^{-2\phi} = e^{-2\rho} =& -\gl^2 \,x^+ x^- \, . \cr}
\eqn\vac
$$
This is just the `linear dilaton' background of non-critical string
theory.  A change of variables, $x^\pm = \pm e^{\pm u^\pm}$,
makes the metric flat and the dilaton field linear in $u^+{-}u^-$.
A `static' black hole is described by
$$\eqalign{
f_i =& 0 \,,  \cr
e^{-2\phi} = e^{-2\rho} =& {M\over \gl} -\gl^2 \,x^+ x^-\,. \cr}
\eqn\static
$$
The Kruskal diagram of this two-dimensional spacetime is shown in
figure 1.  Finally one can describe the metric and dilaton fields
due to an infalling shell of massless matter by patching together
the vacuum solution and a black hole solution across some light-like
line, $x^+ = x_0^+$, (see figure 2),
$$
e^{-2\phi}=e^{-2\rho}=
\cases{\phantom{33333} -\gl^2 x^+x^- &if $x^+<x^+_0$; \cr
-{M\over \gl x^+_0}(x^+-x^+_0) -\gl^2 x^+x^- &if $x^+>x^+_0$. \cr}
\eqn\infal
$$
Note that $\phi$ and $\rho$ are continuous across the matching line.

Null-curves are lines of constant $x^+$ or $x^-$ in our light-cone
coordinates.  The singularity in figure 2 asymptotically approaches
$x^- = - {M\over \lambda^3 x^+_0}$.
Time-like observers with
$x^- > - {M\over \lambda^3 x^+_0}$
cannot escape the singularity, so this line is the global event
horizon formed in classical gravitational collapse.  In the quantum
theory where the black hole evaporates it is useful to introduce the
local notion of a two-dimensional
apparent horizon.  In the classical black hole
solution the event horizon is on a curve where the gradient of the
dilaton field goes from being space-like to time-like.  Recall the
analogy mentioned above between $e^{-\phi}$ and the radius of a
transverse sphere in dimensionally reduced four-dimensional gravity.
A region in space, at a given time, where $\nabla \phi$ is time-like
corresponds to a trapped region in the higher dimensional theory.
In the two-dimensional theory we define the apparent horizon of a
black hole to be the boundary of that region, {\it i.e.} where
$(\nabla \phi)^2=0$.  This way, the apparent horizon will coincide with
the event horizon of a static solution, and its definition corresponds
directly to the standard one in four-dimensional relativity [\wald].

An important feature of this theory is that the strength of gravitational
quantum corrections is ruled by the magnitude of the dilaton field,
$$
g \sim e^{\phi} \, ,
\eqn\coupling
$$
as is evident from the $e^{-2\phi}$ prefactor in front of the gravity
terms in the Lagrangian \lagr .  The coupling strength depends
on position in a black hole background, the theory is free
asymptotically far away and becomes strongly coupled as the classical
singularity is approached.  Note that $g^2 \sim {\gl\over M}$ at
the horizon of a black hole, so quantum corrections are under good
control there while the mass remains large.

\vskip .2 in
\noindent III.  \  \
Hawking radiation is a quantum effect which appears when the matter
fields are quantized in a classical background geometry.  At the
classical level the $f_i$ satisfy free wave equations \cgeom\ and do
not couple to the gravitational degrees of freedom at all.  In the
quantum theory they do couple because of the conformal anomaly.
At the one-loop level the effect of the anomaly can be summarized
in conformal gauge by
adding the well known Liouville term to the gravitational action
[\pol].  Callan {\it et al.} pointed out that the resulting effective
theory could be used not only to compute the Hawking radiation from
a two-dimensional black hole, but also the back-reaction on the
geometry [\cghs].

The quantization of
the gravitational sector of this theory involves a number of
subtle issues (including the contribution to Hawking radiation
of reparametrization ghosts)
which we will not address in this paper.  At the one-loop level,
the contribution due to the conformal anomaly of the matter fields
is the only quantum correction which scales with $N$, and we can
ignore other one-loop terms if the model has a large number of matter
fields.

When the anomaly corrections are taken into account
the dilaton equation in \cgeom\ remains unchanged and the right hand
side of the equation for $\rho$ becomes
${N\over 24}e^{2\phi}\rhopm$.
Particularly convenient combinations of the equations of motion,
for what follows, are given by:
$$\eqalign{
\phipm =&\, (1{-}{N\over 24}e^{2\phi}) \, \rhopm \, ,   \cr
  2\, (1{-}{N\over 12}e^{2\phi}) \,\phipm =&\,
  (1{-}{N\over 24}e^{2\phi}) \,
  (4\phip \phim + \lambda^2 e^{2\rho}) \, .  \cr}
\eqn\eom
$$
In the absence of matter the constraint equations, including Liouville
terms, are
$$\eqalign{
(\p_+^2\phi -2 \rhop\phip ) =&  \,{N\over 24} e^{2\phi}
\bigl(\p_+^2\rho -\rhop\rhop - t_+(x^+) \bigr) \, , \cr
(\p_-^2\phi -2 \rhom\phim ) =&  \,{N\over 24} e^{2\phi}
\bigl(\p_-^2\rho -\rhom\rhom - t_-(x^-) \bigr) \, , \cr}
\eqn\nconstr
$$
where the functions $t_\pm(x^\pm)$ are determined by asymptotic
physical boundary conditions [\cghs].  The quantum corrections in
all these equations are in the $N e^{2\phi}$ terms and they are
small far away from the black hole singularity, as expected.  It is
also easily checked that the linear dilaton vacuum \vac\ remains a
solution of these one-loop corrected equations.

The general solution of the above system of coupled, non-linear,
partial differential equations has not been written in closed form,
to our knowledge.  Nevertheless one can obtain some exact
results using simple manipulations of the equations.  Let us first
establish the existence of a singularity in gravitational collapse.

An infalling shell of matter can be described, much the same way as
in the classical theory, by patching together a vacuum configuration
on the inside and a non-trivial solution of the CGHS-equations
on the outside.  The difference
is that now we do not have the exterior solution in closed form.
However, it will be enough, for our present purposes, to know that
the appropriate solution approaches the classical one
asymptotically far away, where the coupling is weak, and that it
can be matched continuously onto the vacuum across $x^+=x^+_0$.

Consider the equations of motion along a light-like line
infinitesimally above the matter trajectory, $x^+=x^+_0$.
On this line they are ordinary differential equations, in the
variable $x^-$, for the quantities $\Sigma \equiv \phip$ and
$\Xi \equiv \rhop$, with functional coefficients involving
$\phi(x^+_0,x^-)=\rho(x^+_0,x^-)
   = -{1\over 2}\log{(-\lambda^2  x^+_0 x^-)}$.
The bottom equation in \eom\ becomes:
$$
2 (1-{N\over 12}e^{2\phi}) \, \p_-\Sigma
-4(1-{N\over 24}e^{2\phi})\phim \> \Sigma
 = \, (1-{N\over 24}e^{2\phi}) \, \lambda^2 e^{2\rho} \,.
\eqn\sigmaeq
$$
The general solution of this first order equation can be found,
by straightforward integration, to be
$$\eqalign{
\Sigma(x^-) =& - {1\over 2 x^+_0}
         + {K e^{2\phi} \over \sqrt{1{-}{N\over 12}e^{2\phi}}} \cr
            =& - {1\over 2 x^+_0}
         + {K \over {\sqrt{-\gl^2 x^+_0 x^-}
   \sqrt{-\gl^2 x^+_0 x^- - {N\over 12}}}} \, .    \cr}
\eqn\sol
$$
The integration constant is fixed to be
$K={M\over 2 \lambda x^+_0}$ by
the condition that $\Sigma$ approach $\phip$
of the classical black hole solution as $x^- \rightarrow -\infty$.
We are thus able to solve the equations exactly on the collapse
trajectory.  The same technique can in principle be used to calculate
all $x^+$-derivatives of $\phi$ on this line, giving a Taylor
expansion of $\phi$ around it.

The first thing to observe is that $\Sigma$ is singular at
$x^- = - {N\over 12 \lambda^2 x^+_0} \equiv x^-_0$.
The curvature scalar in the
conformal gauge \cgauge\ is given by
$$
R = 8 e^{-2\rho} \rhopm \,.
\eqn\curv
$$
Inserting $\Sigma$ from \sol\ into the first equation of motion in
\eom\ immediately shows that there is in fact a curvature singularity
at $x^- = x^-_0$.
Note that this singularity exists for any choice of dilaton potential
in \lagr .  This is because the dilaton potential enters in the
inhomogeneous
term in the differential equation for $\Sigma$ \sigmaeq , and thus
affects only the $K$-independent term in \sol .

Notice that $\Sigma$ develops an imaginary part when $x^->x^-_0$.
The singularity we have found on the matter trajectory extends into
the region $x^+ > x^+_0$, where it separates real and complex valued
phases of $\phi$.  The curve of singularity is where
${N\over 12}e^{2\phi} = 1$ because the coefficient of the second
order term in the bottom equation in \eom\ vanishes there.
It follows that no real valued solution
in the `Liouville region', where ${N\over 12}e^{2\phi} >1$,
connects to a physical configuration outside
the singularity.

By our definition, an apparent horizon forms on the matter trajectory
when $\Sigma(x^-) = 0$.  This occurs at the roots of the following
quadratic equation,
$$x^- \, (\gl^2 x^+_0 x^- + {N\over 12}) -
   {M^2 \over \gl^4 x^+_0} = \,0 \,.
\eqn\quadr
$$
One solution is in the unphysical $x^- > 0$ region but
the other one
$$
x^- =- \sqrt{({M\over \gl^3 x^+_0})^2 + ({N\over 24 \gl^2 x^+_0})^2}
       - {N\over 24 \gl^2 x^+_0}
\eqn\hor
$$
is where the apparent horizon forms.  This reduces to
$x^- = - {M\over \gl^3 x^+_0}$ in the classical theory, as it should.

A two-dimensional black hole emits Hawking radiation and one expects
the energy loss to cause the newly formed apparent horizon to recede.
This can be checked as follows.  Parametrize the
line of apparent horizon as $x^-=\hat x^-(x^+)$.  By definition
$\phip$ vanishes along this line,
$$
0 = {d\over dx^+}\, \phip \Bigl\vert _{x^-=\hat x^-}
  = \p_+^2 \phi + ({d\hat x^-\over dx^+})\, \phipm \,.
\eqn\hline
$$
The second equation of motion in \eom\ can therefore be written,
$$
-2\,(1{-}{N\over 12}e^{2\phi})\, \p_
+^2\phi\,({d\hat x^-\over dx^+})^{-1}
= \lambda^2 e^{2\rho} \, (1{-}{N\over 24}e^{2\phi}) \,,
\eqn\xbareq
$$
on the horizon line, and the $++$ constraint equation in \nconstr\
can be used to reexpress this as,
$$
{d\hat x^-\over dx^+} = {N\over 12\lambda^2} e^{2\phi-2\rho}
 {(1{-}{N\over 12}e^{2\phi})\over (1{-}{N\over 24}e^{2\phi})}\,
 \bigl(\rhop \rhop - \p_+^2\rho + t_+(x^+)\bigr)\, .
\eqn\dxmdxp
$$
In the classical theory the right hand side vanishes and the
horizon is a light-like line.  The boundary condition that there
be no incoming radiation, except for the infalling shell of matter
at $x^+=x^+_0$, requires that $t_+(x^+)={1\over 4{x^+}^2}$.
For a black hole of large mass, $M>>N\lambda$, the apparent
horizon forms in the weak coupling region, and
${d\hat x^-\over dx^+}$ is well approximated by using the classical
solution \infal\ for $\phi$ and $\rho$ on the right hand side of
\dxmdxp .  To leading order this gives
$$
{d\hat x^-\over dx^+} = {N\over 48\lambda^2{x^+_0}^2} \,.
\eqn\dhat
$$
This expression is manifestly positive so the apparent horizon does
indeed recede, as expected.  Recall that the mass of a two-dimensional
black hole is given by the value of $e^{-2\phi}$ at the horizon.
Thus the rate of energy loss of the black hole is directly related
to the slope of $\hat x^-$.  A straightforward calculation gives the
rate of energy loss as ${N\lambda^2\over 48}$ in exact agreement with
the asymptotic flux of Hawking radiation found by Callan {\it et al.}
in [\cghs].  In this limit the Hawking temperature is independent of
the mass of the black hole [\cghs,\witten].

We conclude this section with some remarks on the qualitative
behavior of solutions of the CGHS-equations far away from the matter
trajectory.  First note that the curve of singularity at
${N\over 12}e^{2\phi}{=}1$ forms inside the trapped region, where
$\nabla \phi$ is time-like.  It follows that the curve of singularity,
like all curves of constant $\phi$ in this region, is always space-like.

The receding apparent horizon is a time-like curve on which
$\phip =0$ by definition.  By contrast, $\phip$ diverges at
the singularity so the apparent horizon and the singularity
curve are unlikely to cross each other.  The behavior that seems to
emerge is that the singularity and the apparent horizon both approach
the same light-like line, from opposite sides, making that line a
global event horizon (see figure 3).  Note that this geometry does
not approach the linear dilaton vacuum smoothly as the incoming
energy goes to zero, because lines of constant $\phi$ are
always time-like in the vacuum.
This suggests that in this theory the final remnant
of an evaporated black hole may carry no mass but nevertheless retains
some `memory' of the initial state.  Quantum coherence would then be
maintained in this theory by light stable remnants carrying
information [\ac,\cghs].  In the real world Planck scale stable
remnants pose serious problems.  Since every possible initial state
must be represented by a distinct remnant, the number of Planck
scale stable particles must be virtually infinite.  These objects
would appear in quantum loops in the scattering of ordinary matter
leading to divergent cross sections.

\vskip .2 in
\noindent IV. \ \
Two main conclusions should be drawn from this paper.  The first
is that physical solutions of the CGHS-equations cannot be continued
from large distances into the Liouville region, where
${N\over 12}e^{2\phi}>1$.  Thus taking the $N\rightarrow\infty$ limit,
without scaling appropriately the distances in the problem with $N$,
cannot serve as a basis for approximating black hole physics in a
systematic way.  However, we have shown that the quantum corrections
of CGHS provide a sensible description of the quantum back reaction
on a large mass ($M>>\lambda N$) black hole.
Our approximations lose their validity towards the end of the black
hole evaporation, when the apparent horizon approaches the singularity
at ${N\over 12}e^{2\phi}=1$, where higher order quantum corrections
become important.  We do not know whether the
singularity persists in a more complete treatment of the quantum
theory.

Finally, we would like to argue that the most interesting issues,
associated with black hole evaporation, have to do with just that
part of the evolution where the CGHS-equations are valid. The
question of information loss is most critical for very large black
holes.  The number of bits of information that can be swallowed in
the formation of a black hole is exponential in its mass in two
spacetime dimensions [\witten] (and exponential in $M^2$ in four
dimensions).
In order to preserve quantum coherence, without invoking stable
remnants, the information contained
in the original black hole must somehow be radiated away during
evaporation because a black hole of lower mass has exponentially
fewer states.  Furthermore, the overwhelming bulk of the information
must be emitted during the time in which the black hole
has $M>N\lambda $.  Unfortunately these equations do not tell us
how this might occur, but we emphasize that
one has to deal with the issue
of quantum information loss long before entering the region of the
singularity.

\noindent
\undertext{Acknowledgement:}
We would like to thank S.~Ben-Menahem, B.~Birnir, C.~Callan,
W.~Fischler, J.~Harvey, R.~Kallosh, J.~Polchinski and A.~Tseytlin
for useful discussions.  This work was supported in part by
NSF grant PHY89-17438.
J.~R. wishes to thank INFN for financial support.

\noindent
\undertext{Note added:}
After completing this paper we were informed by T.~Banks that he
and his collaborators have reached similar conclusions about the
existence of singularities in solutions of the
CGHS-equations.
\refout
\end